\RequirePackage{ifpdf}
\ifpdf 
\documentclass[pdftex]{sigma}
\else
\documentclass{sigma}
\fi

\begin{document}


\renewcommand{\PaperNumber}{123}

\FirstPageHeading

\renewcommand{\thefootnote}{$\star$}

\ShortArticleName{Integrability and Target Space Dif\/feomorphisms}

\ArticleName{Integrability and Dif\/feomorphisms on Target Space\footnote{This
paper is a contribution to the Proceedings of the Seventh
International Conference ``Symmetry in Nonlinear Mathematical
Physics'' (June 24--30, 2007, Kyiv, Ukraine). The full collection
is available at
\href{http://www.emis.de/journals/SIGMA/symmetry2007.html}{http://www.emis.de/journals/SIGMA/symmetry2007.html}}}

\Author{Christoph ADAM~$^\dag$, Joaquin SANCHEZ-GUILLEN~$^\dag$
and Andrzej WERESZCZYNSKI~$^\ddag$}

\AuthorNameForHeading{C. Adam, J. Sanchez-Guillen and A. Wereszczynski}

\Address{$^\dag$~Department of Particle Physics, University of Santiago de
Compostela, Spain} 

\EmailD{\href{mailto:adam@fpaxp1.usc.es}{adam@fpaxp1.usc.es}, \href{mailto:jsanchezguillen@gmail.com}{jsanchezguillen@gmail.com}}

\Address{$^\ddag$~Institute of Physics, Jagellonian University, Reymonta 4,
30-059 Krakow, Poland}

\EmailD{\href{mailto:wereszczynski@th.if.uj.edu.pl}{wereszczynski@th.if.uj.edu.pl}}

\ArticleDates{Received October 18, 2007, in f\/inal form December 10, 2007; Published online December 20, 2007}

\Abstract{We brief\/ly review the concepts of generalized zero curvature conditions and
integrability in higher dimensions, where integrability in this context is
related to the existence of inf\/initely many conservation laws.
Under certain assumptions, it turns out that these conservation laws are, in
fact, generated by a class of geometric target space transformations, namely
the volume-preserving dif\/feomorphisms.
We classify the possible conservation laws of f\/ield theories for the case of
a three-dimensional target space. Further, we discuss some explicit examples.}


\Classification{ 37K05; 37K30; 37K40; 58E50; 81R12; 81T10}

\section{Introduction}\label{sec1}

Integrability, that is the existence of inf\/initely many conservation laws, has
been an invaluable concept for the analysis of nonlinear f\/ield theories in
1+1 dimensions.
In higher dimensions, on the other hand,
much less is known about nonlinear f\/ield theories.
A general concept of integrability has not yet been developed there.

One proposal for integrability in higher dimensions was provided in
\cite{AFSG}, where the zero curvature representation of Shabat and Zakharov
has been generalized to higher dimensions.
Further, it was demonstrated in that paper
that this proposal leads to nonlinear f\/ield theories which have
either inf\/initely many conservation laws in the full theory, or which contain
integrable subsectors, def\/ined by some additional constraint equations on the
f\/ields, such that the solutions belonging to these subsectors
have inf\/initely many conservation laws. This zero curvature representation,
therefore, realizes the concept of integrability in higher-dimensional
non-linear f\/ield theories in a specif\/ic and well-def\/ined manner.

These methods have later been applied to specif\/ic models and to the
analytic construction of both static and time-dependent solutions. For
models with inf\/initely many conservation laws (the so-called AFZ model
and related models), static and time-dependent
solutions have been constructed, e.g., in \cite{AFZ1,AFZ2,Fer1,Wer1}, and in \cite{Fer2,Fer3}, respectively.
Solutions in integrable subsectors of models which are, themselves, not
integrable, have been constructed, e.g., in
\cite{Nic,Wer2,Wer3,ASGVW}
(the Nicole model and versions thereof) and in \cite{Ward1,ASGW-S3}
(diverse models on base space $S^3$). All these models share the property that
their target space has dimension two.

A well-known nonlinear f\/ield theory with three-dimensional target space is
the Skyrme mo\-del~\cite{Skyrme,Skyrme2} with target space SU(2) (or equivalently
the three-sphere $S^3$). Further, this model
contains an integrable subsector, and
the simplest Skyrmion (i.e., the
simplest soliton of the Skyrme model
with baryon number equal to one) belongs to this
integrable subsector, see \cite{FSG1}.

In many cases, it turns out that most of the
new conserved currents
in models and their subsectors are Noether currents
and generalizations thereof, i.e.,
they are related to
 transformations of the target space variables (see \cite{BF1}). So a
direct, geometric approach has been successfully undertaken to f\/ind
those currents, f\/irst for models with two-dimensional target spaces~\cite{gen-int,ab-diff}, and later also for three-dimensional
target spaces~\cite{vol-pres}.

It is the purpose of this review to give a short overview of some of these
recent results on higher-dimensional integrability mentioned above, and to
present some applications. Concretely,
in Section~\ref{sec2} we brief\/ly review the
generalized curvature condition which was proposed in~\cite{AFSG}
as a~possible way to generalize integrability to higher dimensions.
In Section~\ref{sec3} we introduce volume-preserving dif\/feomorphisms on target space.
For the case of three-dimensional target space we then classify for a wide
class of Lagrangians all possible conservation laws, where the conserved
currents in all cases are Noether currents of the volume-preserving
target space dif\/feomorphisms. This section closely follows~\cite{vol-pres}, but provides a slightly more ref\/ined classif\/ication, which
turns out to be useful for applications.
In Section~\ref{sec4} we study as an explicit example the Abelian projection of
Yang--Mills dilaton theory, which turns out to be integrable, and where an
inf\/inite number of analytic static solutions exist. Here we closely
follow the results of~\cite{YM-dil}. Section~\ref{sec5} contains a brief discussion.
In the appendix we provide the calculational details of the classif\/ication
of Section~\ref{sec3}.

\section{Generalized zero curvature condition}\label{sec2}

Here we brief\/ly review the proposal for generalized integrability of~\cite{AFSG}, to which we refer for the details. The structure needed
consists of
a reducible Lie algebra $\tilde {\cal G}$ which is a direct sum of another
Lie algebra ${\cal G}$ and an Abelian ideal ${\cal H}$,
\begin{gather*}
\tilde {\cal G} = {\cal G} \, \oplus \, {\cal H}
\end{gather*}
together with a f\/lat connection
\begin{gather}  \label{1}
A_\mu \in {\cal G}, \qquad
\partial_\mu A_\nu - \partial_\nu A_\mu + [A_\mu ,A_\nu ] =0,
\end{gather}
 and a covariantly constant
vector f\/ield
\begin{gather} \label{2}
B_\mu \in {\cal H}  , \qquad \partial_\mu B^\mu + [ A_\mu ,B^\mu ]=0
\end{gather}
with
  \begin{gather*}
A_\mu = A_\mu^a T^a  , \qquad B_\mu = B_\mu^\alpha P^\alpha , \qquad
[P^\alpha ,P^\beta ]=0 ,
\end{gather*}
where $T^a$ and $P^\alpha$ form a basis in ${\cal G}$ and
${\cal H}$, respectively.
Further, $\mu =1, \ldots, d$ are base space indices.

To gain some intuition, let us f\/irst remark that equation~(\ref{1}) is just a
generalization to higher dimensions of the zero curvature condition of
Zakharov and Shabat in 1+1 dimensions. Further, equation~\eqref{2} in some sense
just generalizes the Lax pair $\dot L = [L,M]$ to higher dimensions.
Another important point is that equations~(\ref{1}), (\ref{2}) are not chosen
arbitrarily but may, in fact be derived from a generalized curvature condition
as follows.

Firstly, the zero curvature condition equation~(\ref{1}) of Zakharov and Shabat
may be derived as a consequence of the path independence
of the Wilson line (or parallel transport) operator
\begin{gather*} 
W=P\exp \left( \int_0^\sigma d\sigma' A_\mu \frac{dx^\mu}{d\sigma'}\right),
\end{gather*}
where $P$ indicates the path ordering.
In an analogous way,  equations~(\ref{1}), (\ref{2}) may be derived from the
 hypersurface independence of the following
hypersurface ordered operator $V$ in $d$ dimensions,
\begin{gather*} 
V=\tilde P \exp \left( \int_{\Sigma_{d-1}} d\sigma^1 \cdots d \sigma^{d-1}
W^{-1} \tilde B_{\mu_1 \ldots \mu_{d-1}} W \frac{dx^{\mu_1}}{d\sigma^1}
\cdots \frac{dx^{\mu_{d-1}}}{d\sigma^{d-1}} \right) .
\end{gather*}
Here the $d-1$ form $\tilde B$ is the Hodge dual of the vector (one-form)
of equation~(\ref{2})
(in fact, $\tilde B$~is the more natural object from the point of
view of generalized integrability). Further,  $\Sigma_{d-1}$ is a based,
ordered, closed hypersurface with base point $x_0 \equiv x$ $(\sigma_j=0)$.
$\tilde P$ is the hypersurface ordering, which we shall explain a bit more in
a moment. So far this is a generalization of the non-Abelian Stokes calculus
formulated in three dimensions by Schlesinger in 1927 \cite{Schl}.

The hypersurface independence of $V$, in turn, may be derived from the zero
curvature condition for a connection ${\cal A}$ in higher loop space
$\Omega^n (M,x_0)$ where
\begin{gather*}
\Omega^{n}(M,x_0 ) = \{ \gamma : S^n \to M , \gamma (0,\ldots ,0)=x_0\}
\end{gather*}
Explicitly, the connection ${\cal A}$ reads
\begin{gather} \label{a}
{\cal A}
= \int_{\Sigma_{d-1}} d\sigma^1 \cdots d \sigma^{d-2}
W^{-1} \tilde B_{\mu_1 \ldots \mu_{d-1}} W \frac{dx^{\mu_1}}{d\sigma^1}
\cdots \frac{dx^{\mu_{d-2}}}{d\sigma^{d-2}} \delta x^{\mu_{d-1}},
\end{gather}
where $\delta x^{\mu_{d-1}}$ is the dif\/ferential on higher
loop space which provides
an arbitrary inf\/initesimal variation of the higher loop.
A closed ordered based hypersurface $\Sigma^{d-1}$ may be interpreted as a
closed loop in loop space $\Omega^{d-2}$, and this observation allows to
understand the hypersurface ordering. It is just ordinary path ordering of
the corresponding ordinary loop in higher loop space $\Omega^{d-2}$.

We want to emphasize that the conditions equations~(\ref{1}), (\ref{2}) are
{\em sufficient, local} conditions for the zero curvature condition on the
connection  ${\cal A}$ of equation~(\ref{a}), but certainly they are not the most
general ones.

After this brief review of the generalized zero curvature condition, we
assume that equations~(\ref{1}), (\ref{2}) hold and make the following additional
simplifying assumptions that
\begin{itemize}\itemsep=0pt
\item ${\cal G}$ is a semisimple Lie algebra (e.g.~su(2)) with
\begin{gather} \label{ass1}
\quad [T^a ,T^b] = f^{ab}_c T^c
\end{gather}
and structure constants $f^{ab}_c$.
\item
  ${\cal H}$ is a (in general, reducible) representation space of ${\cal G}$:
\begin{gather}
[T^a ,P^\alpha ] =R^{\alpha\beta}
(T^a) P^\beta.
\end{gather}
\item $A_\mu$ is explicitly f\/lat:
\begin{gather}
A_\mu = g^{-1} \partial_\mu g  , \qquad
g\in   G
\end{gather}
(where e.g.\ $  G =$ SU(2)) such that only equation~(\ref{2}) provides a
nontrivial condition ($D_\mu B^\mu =0$).
\item Under these conditions, the currents
\begin{gather} \label{ass4}
J_\mu = gB_\mu g^{-1}
\end{gather}
are automatically conserved,
$\partial^\mu J_\mu =0$, and therefore the number of the conserved currents
equals the dimension of the representation space ${\cal H}$,
dim ${\cal H}$. If
dim ${\cal H} = \infty$,  then we say that the corresponding f\/ield theory is
integrable.
\end{itemize}

In general, the conserved currents  of an integrable theory
may be either Noether currents or
may be related to hidden symmetries.
Under the assumptions equations~(\ref{ass1})--(\ref{ass4}), however, the currents
$J_\mu$ turn out to be Noether currents
of geometric target space transformations, where the target space is spanned by
the parameters of $g\in  G$.

A f\/irst example for this structure is like follows.
\begin{itemize}\itemsep=0pt
\item The Lie group $ G$ is  SU(2) where, however, its elements
$g\in  G$
are restricted to the equator of SU(2) such that the target space is
two-dimensional.
\item The representation space ${\cal H}$
is the space of representations of SU(2)
  with arbitrary integer angular momentum quantum number $l$, but magnetic
quantum number $m$ restricted to~$\pm 1$,
\begin{gather*}
{\cal H}=\{ {\rm  reps.} \,\,  R_{lm} \,\,  {\rm of \, SU(2)}, \
m=\pm 1 , \ l=1, \ldots   ,\infty\}.
\end{gather*}
\item Then the conserved currents $J_\mu$ of equation~(\ref{ass4}) turn out to
generate area preserving dif\/feomorphisms on the two-dimensional
  target space (which may be, e.g., the two-sphe\-re~$S^2$,
but this depends on the Lagrangian).
\end{itemize}
A detailed discussion of this case may be found in \cite{BF1},
or in \cite{gen-int,ab-diff}.

Another class of theories with three-dimensional target spaces is obtained
when
the group element $g$ is assumed to take values in the full unrestricted
group SU(2). There it turns out that the resulting conservation laws are
generated by some subsets of the generators of volume-preserving
dif\/feomorphisms on that target space.
This case is discussed in the next section, where also a
classif\/ication of the conservation laws of these theories is given.
More details may be found in~\cite{vol-pres}.

\section{Conservation laws for Skyrme-type models}\label{sec3}

\subsection[Volume-preserving diffeomorphisms]{Volume-preserving dif\/feomorphisms}

Let us start with a three-dimensional manifold (later to be identif\/ied with
target space) with local coordinates $ X^i$ and with a volume form which in
local coordinates reads
\begin{gather*}
dV = h(X^i) dX^1 \wedge dX^2 \wedge dX^3,
\end{gather*}
where $h$ is the volume density. Further, a dif\/feomorphism is
an inf\/initesimal transformation
\begin{gather*}
X^i \to X^i + \epsilon Y^i (X^j),
\end{gather*}
where $\epsilon$ is inf\/initesimal, and the $Y^i$ are arbitrary functions of
the $X^j$.
A volume-preserving dif\/feomorphism has to obey
\begin{gather*}
\partial_i (hY^i)\equiv \frac{\partial}{\partial X^i} (hY^i) =0,
\end{gather*}
in addition. The corresponding vector f\/ields
\begin{gather*}
v^{(Y)} = Y^i\partial_i
\end{gather*}
form a closed Lie algebra, that is
\begin{gather*}
[v^{(Y)} ,v^{(\tilde Y)} ] = v^{(\tilde{\tilde Y}   )}
\end{gather*}
such that
\begin{gather*}
 \tilde{\tilde Y}^i =(\partial_j Y^i) \tilde Y^j -
(\partial_j \tilde Y^i )Y^j   , \qquad \partial_i (h\tilde{\tilde Y}^i)=0
\end{gather*}
is again a volume-preserving dif\/feomorphism.
For later convenience we change coordinates according to
\begin{gather*}
u\equiv X^1 +iX^2 ,\qquad \xi \equiv X^3, \nonumber \\
Y^u \equiv Y^1 +iY^2 ,\qquad Y^\xi \equiv Y^3 .
\end{gather*}
These new coordinates are
especially useful for a parametrization of $g\in$ SU(2),
\begin{gather} \label{g}
g=\exp (i \xi \vec n \cdot \vec \sigma ) = \cos \xi +i\sin \xi \, \vec n \cdot
\vec \sigma
\end{gather}
provided that we also replace the unit vector f\/ield $\vec n$ by a
complex f\/ield $u$ via
stereographic projection
\begin{gather} \label{ster-proj}
\vec n \rightarrow u=\frac{n_1 +in_2}{1+n_3} .
\end{gather}
Further, we assume from now on the following form of the volume density
$h=h(u\bar u ,\xi )$ for simplicity.
Finally, we interpret $u$, $\xi$  as target space variables of a Lagrangian f\/ield
theory with general Lagrangian
${\cal L}(u,\bar u,\xi ,u_\mu ,\bar u_\mu ,\xi_\mu) $, where $  u_\mu \equiv
\partial_\mu  u$, etc.,
then the Noether currents corresponding to the vector f\/ields
$v^{(Y)} $ generating volume-preserving dif\/feomorphisms on target space
are given by
\begin{gather} \label{dif-cur}
J^{(Y)}_\mu = Y^u \Pi_\mu + Y^{\bar u} \bar \Pi_\mu +Y^\xi P_\mu
\end{gather}
with the usual canonical four-momenta
\begin{gather*}
\Pi_\mu \equiv \partial_{u^\mu} {\cal L}  ,\qquad
P_\mu \equiv \partial_{\xi^\mu} {\cal L} .
\end{gather*}
The charges $Q^{(Y)} =\int d^3 {\bf r} J^{(Y)}_0$ generate a Lie algebra
isomorphic to the algebra of the vector f\/ields  $v^{(Y)} $
via the Poisson bracket, as usual.

\subsection[Classification of conserved currents]{Classif\/ication of conserved currents}

We now specialize to the class of Lagrangians
\begin{gather} \label{lag-abcde}
{\cal L}(a,b,c,\xi ,d,e),
\end{gather}
where
\begin{gather*}
a=u\bar u  ,  \qquad b=u^\mu \bar u_\mu  ,  \qquad d=\xi^\mu \xi_\mu,
\qquad
c=(u^\mu \bar u_\mu)^2 - u_\mu^2 \bar u_\nu^2  ,
\qquad e=\xi^\mu u_\mu \xi^\nu \bar u_\nu .
\end{gather*}
To motivate this choice let us mention that, e.g., the Skyrme model belongs to
this class. Indeed, the Skyrme model has the Lagrangian
${\cal L}_{\rm Sk}=\frac{m^2}{2} {\cal L}_2 -
\lambda {\cal L}_4 $ where
\begin{gather}
{\cal L}_2 = \mbox{tr} (g^{-1}g_\mu g^{-1} g^\mu ) = d+4b
\frac{\sin^2 \xi}{(1+a)^2}, \nonumber \\
{\cal L}_4 = \mbox{tr} [g^{-1}g_\mu , g^{-1} g_\nu ]^2 =
\frac{\sin^2 \xi}{(1+a)^2}
(bd - e) + \frac{\sin^4 \xi}{(1+a)^4}c \nonumber
\end{gather}
and $g$ is the SU(2) group element of equation~(\ref{g}).

For the class of Lagrangians (\ref{lag-abcde}) we now want to f\/ind which
subsets of the currents (\ref{dif-cur}) are conserved under which conditions.
The calculation is lengthy and is, therefore, relegated to the Appendix.
Here, we present the resulting
classif\/ication in the following four tables. In Table~\ref{table1} the f\/ields just
obey the f\/ield equations and, therefore, there is a one to one correspondence
between symmetries and conservation laws. In Tables~\ref{table2}--\ref{table4}, on the other hand,
the f\/ields have to obey certain f\/irst order equations (``integrability
conditions''), in addition. These integrability conditions are not of the
Euler--Lagrange type and, therefore, there is no longer a one to one
correspondence between symmetries and conservation laws, see~\cite{syms}
for a detailed discussion.

\begin{table}[t] {\centering
\caption{}\label{table1}
\vspace{1mm}

\begin{tabular}{|rl|}
\hline
\multicolumn{2}{|c|}{{\tsep{1ex} \bf No integrability conditions\bsep{0.5ex} }} \\
\hline
 a) & no condition on ${\cal L}$.  \\
& Generically there exists only one vector f\/ield $Y$: \\
& $Y^u = iu$, \, $Y^{\bar u} =-i\bar u $, \, $Y^\xi =0$. \\
\hline
b) & ${\cal L}=
{\cal F}(hb,h^2 c,d ,he)$. \\
& There exist f\/initely many $Y$ generating the \\
&  isometries of the target space metric. \\
\hline
c) & ${\cal L}_b =0$ and ${\cal L}_e =0$.  \\
& $Y$ form the Abelian subalgebra ($\tilde G=\tilde G(a)$): \\
& $Y^u  =iu\tilde G_a$, \, $ Y^{\bar u} =-i\bar u \tilde G_a $,
\, $ Y^\xi =0$. \\
\hline
d) & conditions b) and c) on ${\cal L}$, \\
& and factorizing $h= h_1 (a)h_2 (\xi)$. \\
& $Y$ forms the non-Abelian subalgebra (for $G=G(u,\bar u)$): \\
& $Y^u = ih_1^{-1}G_{\bar u}$, \, $Y^{\bar u} =-ih_1^{-1} G_u$, \,
$Y^\xi =0$. \\
\hline
\end{tabular}

}

\medskip

\noindent {\small Case a) corresponds to the symmetry $u \to e^{i\alpha} u$.
Case b) implies that the Lagrangian can be expressed by the pullback of
a certain target space metric, such that $h$ is the Riemannian volume density
of that metric. One example for this case is the Skyrme model.
One example for case d) is provided by the Abelian projection of Yang--Mills
dilaton theory, which is discussed in Section~\ref{sec4}.}
\end{table}


\begin{table}[th]{\centering
\caption{}\label{table2}

\vspace{1mm}

\begin{tabular}{|rl|}
\hline
\multicolumn{2}{|c|}{\tsep{1ex}{\bf Integrability condition
$u^2 \bar u_\mu^2 - \bar u^2 u_\mu^2 =0$}\bsep{0.5ex} } \\
\hline
a) &   ${\cal L}_b =0$ and ${\cal L}_e =0$.  \\
& $Y$ forms the Abelian subalgebra (for $G=G(a,\xi)$): \\
& $Y^u = ih^{-1}uG_a$, \, $Y^{\bar u} =-ih^{-1}\bar u G_a$, \, $Y^\xi =0$. \\
\hline
b) & ${\cal L}_e =0$.  \\
& $Y$ form the Abelian subalgebra ($\tilde G=\tilde G(a)$): \\
& $Y^u  =iu\tilde G_a$, \, $ Y^{\bar u} =-i\bar u \tilde G_a $,
\, $ Y^\xi =0$. \\

\hline
\end{tabular}

}

\medskip

\noindent
{\small The integrability condition $u^2 \bar u_\mu^2 - \bar u^2 u_\mu^2 =0$
may also be expressed like $\partial^\mu ({\rm mod}(u)) \partial_\mu
({\rm arg}(u))=0$, which provides a more geometric interpretation.}
\end{table}

\begin{table}[th]{\centering
\caption{}\label{table3}

\vspace{1mm}

\begin{tabular}{|rl|}
\hline
\multicolumn{2}{|c|}{\tsep{1ex}{\bf Integrability conditions $u^\mu \xi_\mu =0$}\bsep{0.5ex}} \\
\hline
 a) & \tsep{0.5ex} no condition on ${\cal L}$; or ${\cal L}=
{\cal F}(hb,h^2c,d ,he)$.  \\
& $Y$ forms the Abelian subalgebra (for $G=G(a,\xi)$): \\
& $Y^u = ih^{-1}uG_a$, \, $Y^{\bar u} =-ih^{-1}\bar u G_a$, \, $Y^\xi =0$. \\
& {\em And} the further integrability condition
 $u^2 \bar u_\mu^2 - \bar u^2 u_\mu^2 $ holds. \\
\hline
b) & ${\cal L}_b =0$. \\
& $Y$ forms the Abelian subalgebra (for $G=G(a,\xi)$): \\
& $Y^u = ih^{-1}uG_a$, \, $Y^{\bar u} =-ih^{-1}\bar u G_a$, \, $Y^\xi =0$. \\
\hline
c) & \tsep{0.5ex} ${\cal L}_b =0$ and ${\cal L}=
{\cal F}(hb,h^2c,d ,he)$. \\
& $Y$ form the subset $Y^\xi_\xi =0$ (is {\em not}  a subalgebra). \\
\hline
d) &  \tsep{0.5ex} ${\cal L}_b =0$ and ${\cal L}=
{\cal F}(hb,h^2c,d ,he)$
and ${\cal W}({\cal L})  =0$. \\
 & no further condition on $Y$.   \\
\hline
\end{tabular}

}

\medskip

\noindent
{\small Case a) is obeyed by many conf\/igurations of the Skyrme model.
E.g., the simplest Skyrmion with baryon number one as well as many ans\"atze for
Skyrmion conf\/igurations satisfy the conditions of case~a).
Case~d): the ``weight number''  ${\cal W}$ is def\/ined for monomials
of f\/irst derivatives of f\/ields as
${\cal W}= \mbox{power}(u_\mu ) +\mbox{power}(\bar u_\mu ) -2
\mbox{power}(\xi_\mu )$, which gives
e.g.~${\cal W}(b)=2$, ${\cal W}(e)=-2$.}

\end{table}

\begin{table}[th]{\centering

\caption{}\label{table4}

\vspace{1mm}

\begin{tabular}{|rl|}
\hline
\multicolumn{2}{|c|}{\tsep{1ex}{\bf Integrability conditions
$u_\mu^2 =0$ and $u^\mu \xi_\mu =0$}\bsep{0.5ex}} \\
\hline
 a) & no condition on ${\cal L}$.  \\
& $Y$ forms the Abelian subalgebra (for $G=G(a,\xi)$): \\
& $Y^u = ih^{-1}uG_a$, \, $Y^{\bar u} =-ih^{-1}\bar u G_a$, \, $Y^\xi =0$. \\
\hline
b) & \tsep{0.5ex}${\cal L}= {\cal F}(hb,h^2c,d ,he)$ \\
& $Y$ form the subset $Y^\xi_\xi =0$ (is {\em not}  a subalgebra). \\
\hline
c) & \tsep{0.5ex} ${\cal L}= {\cal F}(hb,h^2c,d ,he)$ and ${\cal W}({\cal L})   =0$. \\
 & no further condition on $Y$.   \\
\hline
\end{tabular}

}

\medskip

\noindent
{\small Case b) is obeyed by many conf\/igurations of the Skyrme model.
E.g., the simplest Skyrmion with baryon number one or the rational map
ans\"atze for Skyrmion conf\/igurations satisfy the conditions of case b).}

\end{table}

\section{Example: Abelian projection of YM dilaton theory}\label{sec4}

Here we want to demonstrate that the Abelian projection of Yang--Mills dilaton
theory is integrable in our sense.
It belongs, in fact, to case d) of Table~\ref{table1}.
A more detailed discussion can be found in~\cite{YM-dil}.
The Lagrangian of Yang--Mills dilaton theory is
\begin{gather*}
{\cal L}= \frac{1}{4} (2\xi^\mu \xi_\mu -e^{-2\kappa \xi}
F^{a\mu\nu}F^a_{\mu\nu}),
\end{gather*}
where $A^a_\mu$ is an SU(2) Yang--Mills f\/ield and $F^a_{\mu\nu}$ is the
corresponding f\/ield strength. Next, we want to employ the
Cho--Faddeev--Niemi decomposition of the gauge f\/ield,
\begin{gather*}
A^a_\mu  =  n^a C_\mu  +  \epsilon^{abc} n^b_\mu n^c
 +  W^a_\mu ,
\end{gather*}
where $C_\mu$ is an Abelian gauge f\/ield, $n^a$ is a unit vector in color
space, and the so-called ``valence f\/ield'' $W^a_\mu$ is perpendicular to $n^a$
in color space, $n^a W^a_\mu =0$. To be consistent, the decomposition f\/ields
have to obey the constraint
\begin{gather} \label{constr}
\partial^\mu W^a_\mu + C_\mu \epsilon^{abc} n^b W^c_\mu +
n^a W^b_\mu n^b_\mu =0 .
\end{gather}
This constraint makes that the number of degrees of freedom of the original
gauge f\/ield and of the decomposition match, and further, it provides the
correct behaviour under gauge transformations for the decomposition f\/ields,
which inf\/initesimally read
\begin{gather*}
\delta n^a = \epsilon^{abc} n^b \alpha^c, \qquad
\delta W^a_\mu  = \epsilon^{abc} W^b_\mu  \alpha^c, \qquad
\delta C_\mu = n^a \alpha^a_\mu .
\end{gather*}
In a next step, we perform
the Abelian projection, which consists in setting the
valence f\/ield equal to zero,
\begin{gather*}
W^a_\mu =0.
\end{gather*}
Observe that the Abelian projection is gauge invariant and obeys the
constraint (\ref{constr}). The resulting gauge f\/ield
$
\hat A^a_\mu = n^a C_\mu + \epsilon^{abc}n^b_\mu n^c
$
still is a full SU(2) connection, but with Abelian f\/ield strength.
The resulting  Abelian projected Yang--Mills dilaton theory is already
integrable, that is, it has inf\/initely many conserved currents, see~\cite{YM-dil} for details. Here we make the further simplifying
assumption $C_\mu \equiv 0$ (which is no longer gauge invariant).
The resulting Abelian projected Lagrangian is
\begin{gather} \label{lag-AP}
{\cal L}_{\rm AP}   =
\frac{1}{4}  (   2\xi^\mu \xi_\mu   -
e^{-2\kappa \xi} H^{\mu\nu} H_{\mu\nu}  )
\end{gather}
with
\begin{gather*}
H_{\mu \nu} =
\epsilon^{abc}n^a n^b_\mu n^c_\nu
\end{gather*}
or, after the stereographic projection (\ref{ster-proj}),
\begin{gather*}
{\cal L}_{\rm AP} = \frac{1}{2}\xi^\mu \xi_\mu -2 e^{-2\kappa \xi}
\frac{(u^\mu \bar u_\mu )^2 -u^2_\mu \bar u^2_\nu }{(1+u\bar u)^4}
\equiv  \frac{1}{2} d -2h^2 c,
\end{gather*}
where $h=h_1 (a) h_2 (\xi ) \equiv (1+a)^{-2}
e^{-\kappa \xi}$.
It corresponds to case d) of Table~\ref{table1} and has, therefore, inf\/initely many
symmetries and inf\/initely many conservation laws.

We now want to use our explicitly integrable parametrization of the Abelian
projection of Yang--Mills dilaton theory to discuss the problem of static
solutions. For that purpose, we should f\/irst review the known results on that
issue. It is known that there exist static, sphaleron type solutions in
Yang--Mills dilaton theory. For the fully non-Abelian theory, solutions both
for radially and cylindrically symmetric ans\"atze are known numerically,
whereas for the Abelian subsector  solutions
for radially and cylindrically symmetric ans\"atze are known analytically.
In the latter case, it is further known that the energy of the analytic
solutions grows linearly with a certain integer $m$ from the ansatz
(the magnetic quantum number). The latter fact points to the existence of
a Bogomolny bound in the Abelian projection, but an ansatz-independent
derivation of this Bogomolny bound has not yet been given in the literature.

In our integrable Abelian projection of YM dilaton theory, the analytic
solutions may be calculated easily by quadratures, and the Bogomolny bound may
be derived explicitly. Indeed, upon introducing spherical polar coordinates
$(r,\theta ,\varphi )$ in three-dimensional base space, the ansatz $\xi = \xi
(r)$, $u = v(\theta )\exp (im\varphi )$ turns out to be consistent with the
static f\/ield equations because of the base space symmetries of the theory. The
resulting ordinary dif\/ferential equations for $\xi (r)$ and
$v (\theta )$ turn out to be solvable by quadratures, such that the
corresponding exact analytic solutions may be calculated easily.
The solvability by quadratures of the f\/ield equations might be
related to the integrability of the theory.
For details we refer to~\cite{YM-dil}.

Finally, the Bogomolny bound may be derived easily within our parametrization.
Indeed, we f\/ind for the  energy corresponding to the Lagrangian
(\ref{lag-AP}) for static conf\/igurations
\begin{gather*}
E_{\rm AP} = \frac{1}{2}\int d^3 {\bf r} \left( (\nabla \xi )^2 +
e^{-2\kappa \xi }
\vec H^2 \right) =   \frac{1}{2}\int d^3 {\bf r} \left( \nabla \xi  -  e^{-\kappa \xi }
\vec H \right)^2 + \int d^3 {\bf r} e^{-\kappa \xi } \nabla \xi \cdot
\vec H  \nonumber \\
\phantom{E_{\rm AP}} {} \ge  \int d^3 {\bf r} e^{-\kappa \xi } \nabla \xi \cdot \vec H
\equiv E_{\rm Bog.} \nonumber
\end{gather*}
(where $ \vec H  $ is the
Hodge dual of $H_{jk}$)
and, therefore, the Bogomolny equation
\begin{gather*}
 \nabla \xi - e^{-\kappa \xi } \vec H =0.
\end{gather*}
All the analytic static solutions
mentioned above satisfy this equation and are, therefore, Bogomolny solutions.
Further, the Bogomolny energy $E_{\rm Bog.}$ may be expressed by the
winding number  of a map S$^3$ $\to $ S$^3$, see again~\cite{YM-dil}.

To recapitulate, our main results for the
Abelian projection of YM dilaton theory are that
\begin{itemize}\itemsep=0pt
\item there exist inf\/initely many symmetries and inf\/initely many
conserved currents,
\item this fact may explain the inf\/initely many analytic solutions
(this still is a conjecture, which exploits the
  analogy to the lower dimensional cases),
\item there exist both a  Bogomolny bound and a Bogomolny equation for static
conf\/igurations, and the latter is solved by all known analytic solutions.
\end{itemize}

\section{Discussion and outlook}\label{sec5}

It was the purpose of this article to brief\/ly review some recent developments
in the attempts to generalize the concept of integrability to
higher-dimensional nonlinear f\/ield theories. We gave a brief introduction to
the general proposal for higher-dimensional integrabi\-lity of~\cite{AFSG} and then
showed how, under certain additional assumptions, this higher-dimensional
integrabi\-li\-ty is related to certain geometric target space transformations
(concretely, volume preserving dif\/feromorphisms) which provide inf\/initely many
conservation laws. We discussed in some more detail the case of a
three-dimensional target space and, as a specif\/ic example, the Abelian
projection of Yang--Mills dilaton theory and its static analytic solutions.
Some more applications to specif\/ic theories have already been studied
(see, e.g., the references quoted in the Introduction),
which already demonstrates the usefulness and importance of the concept of
generalized integrability for the study of higher-dimensional nonlinear f\/ield
theories.
There exist, however,
many more applications which
are still open to further investigation. One obvious
application is the search for time-dependent solutions (e.g.~$Q$-balls) in
theories where till now only static solutions have been found (e.g. in the
integrable submodel of Yang--Mills dilaton theory of the previous section).
Another possibility for generalizations consists in the choice of larger
groups ${\cal G}$ instead of SU(2) in the integrability construction
discussed in Section~\ref{sec2}. This leads to integrable theo\-ries with
higher-dimensional target spaces. The search for an integrable submodel of
Einstein Yang--Mills dilaton theory and for analytic solutions within this
submodel would be an obvious candidate, especially as for this theory only
numerical solutions are known so far.

The generalizations and further investigations mentioned here still deal
with a connection which is trivially f\/lat, $A_\mu =g^{-1} \partial_\mu g$,
such that the zero curvature condition (\ref{1}) of Section~\ref{sec2}
is trivially fulf\/illed, and equation~(\ref{2}) remains the only nontrivial
generalized zero curvature condition. A further possible generalization
consists in treating equation~(\ref{1}) as a nontrivial condition, too, which
generates nontrivial constraints on the connection.  The resulting
modif\/ied generalized integrability might then lead to
nonlocal conserved currents and to conservation laws which are not generated
by geometric transformations, as is well-known to be the case in 1+1
dimensions. This line of investigation is, however, almost completely
unexplored, and the above remarks are, therefore, mainly a proposal at the
moment.

\appendix

\section{Appendix}\label{appendix}

In this appendix we provide the details of the calculation of the conserved
currents (\ref{dif-cur}) for the class of Lagrangians (\ref{lag-abcde}).
The Lagrangians ${\cal L} =
{\cal L}(a,b,c,\xi ,d,e)$ have the following canonical four-momenta
\begin{gather*}
\Pi_\mu = {\cal L}_{u^\mu} = \bar u_\mu {\cal L}_b +2 (b \bar u_\mu -
\bar u_\nu^2 u_\mu ){\cal L}_c + (\xi^\nu \bar u_\nu) \xi_\mu {\cal L}_e,
\\
P_\mu = {\cal L}_{\xi^\mu} = 2 \xi_\mu {\cal L}_d + ( (\xi^\nu \bar u_\nu)
u_\mu + (\xi^\nu  u_\nu) \bar u_\mu ) {\cal L}_e ,
\end{gather*}
f\/ield equations
\begin{gather*}
\partial^\mu \Pi_\mu = {\cal L}_u = \bar u {\cal L}_a , \qquad
\partial^\mu P_\mu = {\cal L}_\xi ,
\end{gather*}
and we need the following useful identities
\begin{gather*}
u^\mu \Pi_\mu = b{\cal L}_b +2 c{\cal L}_c + e{\cal L}_e, \\
\bar u^\mu \Pi_\mu = \bar u_\mu^2 {\cal L}_b + (\bar u_\mu \xi^\mu )^2
{\cal L}_e, \\
\xi^\mu \Pi_\mu = (\xi^\mu \bar u_\mu) {\cal L}_b +
2 (b \xi^\mu \bar u_\mu - \bar u_\nu^2 \xi^\mu u_\mu) {\cal L}_c +
d \xi^\mu \bar u_\mu {\cal L}_e, \\
u^\mu P_\mu = 2(\xi^\mu u_\mu ) {\cal L}_d + ((\xi^\mu \bar u_\mu ) u_\nu^2 +
b \xi^\mu u_\mu ) {\cal L}_e, \\
\xi^\mu P_\mu = 2 d {\cal L}_d +2 e{\cal L}_e .
\end{gather*}
Now we want to calculate the divergence of the Noether currents
(\ref{dif-cur})
\begin{gather*}
\partial^\mu J^{(Y)}_\mu = (Y^u{}_u u^\mu + Y^u{}_{\bar u} \bar u^\mu
+ Y^u{}_\xi \xi^\mu )\Pi_\mu +
(Y^{\bar u}{}_u u^\mu + Y^{\bar u}{}_{\bar u} \bar u^\mu
+ Y^{\bar u}{}_\xi \xi^\mu )\bar \Pi_\mu  \nonumber \\
\phantom{\partial^\mu J^{(Y)}_\mu =}{}+(Y^\xi{}_u u^\mu + Y^\xi{}_{\bar u} \bar u^\mu
+ Y^\xi{}_\xi \xi^\mu )P_\mu +
 Y^u \partial^\mu \Pi_\mu + Y^{\bar u} \partial^\mu \bar \Pi_\mu +
Y^\xi \partial^\mu P_\mu .
\end{gather*}
After a lengthy but straight-forward calculation, and using
\begin{gather*}
Y^u{}_u + Y^{\bar u}{}_{\bar u} = h[(h^{-1})_u Y^u + (h^{-1})_{\bar u}
Y^{\bar u} + (h^{-1})_\xi Y^\xi ] - Y^\xi {}_\xi
\end{gather*}
(which easily follows from $\partial_i (hY^i )=0$),
 we f\/ind
\begin{gather} \label{app-cons-eq}
\partial^\mu J^{(Y)}_\mu = {\rm I} + {\rm II} + {\rm III},
\end{gather}
where
\begin{gather}
{\rm I} = h(\bar u Y^u + u Y^{\bar u} ) [(h^{-1})_a u^\mu \Pi_\mu
+ h^{-1} {\cal L}_a ]
 +  hY^\xi [(h^{-1})_\xi u^\mu \Pi_\mu + h^{-1} {\cal L}_\xi ]
\nonumber \\
\phantom{{\rm I} =}{} + Y^\xi {}_\xi (\xi^\mu P_\mu - u^\mu \Pi_\mu ) , \label{app-I}
\end{gather}
and
\begin{gather*}
{\rm II}  =
(Y^u{}_{\bar u} \bar u_\mu^2 + Y^{\bar u} {}_{u} u^2_\mu )
{\cal L}_b .
\end{gather*}
For the special choice
\begin{gather*}
Y^u = i h^{-1} uG_a  ,\qquad Y^{\bar u} = -i h^{-1} \bar uG_a   ,\qquad
Y^\xi =0,\qquad G=G(a,\xi)
\end{gather*}
this simplif\/ies to
\begin{gather*}
{\rm II}   =
[\partial_a (h^{-1} G_a)] (u^2 \bar u^2_\mu - \bar u^2 u^2_\mu )
{\cal L}_b .
\end{gather*}
Further,
\begin{gather}
{\rm III} = (u^\mu \xi_\mu )^2 {\cal L}_e Y^{\bar u}{}_u -
u^2_\nu \bar u^\mu \xi_\mu (2 {\cal L}_c Y^{\bar u}{}_\xi -
{\cal L}_e Y^\xi {}_u ) \nonumber
\\
\phantom{{\rm III} =}{} + u^\mu \xi_\mu [({\cal L}_b +2b {\cal L}_c + d{\cal L}_e )Y^{\bar u}{}_\xi
+ (2{\cal L}_d + b{\cal L}_e )Y^\xi {}_u] + {\rm h.c.} \label{app-III}
\end{gather}
(Note that the above, correct, expression equation~(\ref{app-III}) dif\/fers from the
corresponding one in~\cite{vol-pres} by one sign and one factor of two,
which are incorrect in~\cite{vol-pres}. This small error in that
reference has, however,
absolutely no signif\/icance for the results of~\cite{vol-pres}.)
From these results it is not dif\/f\/icult to reconstruct the classif\/ication
presented in Tables \ref{table1}--\ref{table4} of the main text. Let us focus on Table 1 for the
moment. If no conditions are imposed on the Lagrangian, then the $Y^i$ have to
obey $uY^{\bar u} + \bar u Y^u =0$ and $Y^\xi =0$ together with
$Y^u{}_\xi = 0$, $Y^u{}_{\bar u} =0$, etc, which has the only solution
$Y^u =iu$, $Y^{\bar u} =-i\bar u$, which is just case a) of Table~\ref{table1}.
In case b), the condition ${\cal L} = {\cal F}(hb,h^2 c,d,he)$
implies that the f\/irst two terms in I, equation~(\ref{app-I}), are absent,
but for general ${\cal F}(hb,h^2 c,d,he)$, it is still dif\/f\/icult to
reconstruct the corresponding
target space metric. However, if we restrict to the subclass
of models ${\cal L} ={\cal F}[d+hb, 4h(bd-e) + h^2 c]$ (to which e.g.~the
Skyrme model belongs), then the target space
metric is just
\begin{gather*}
ds^2 = d\xi \otimes d\xi + h \frac{1}{2} (du \otimes d\bar u +
d\bar u \otimes du)
\end{gather*}
and it is not dif\/f\/icult to check that the resulting conditions on $Y^i$ in
that case are just the Killing equations for the above target space metric.
Further, cases c) and d) are easy to check, because in these cases most terms
in the conservation equation (\ref{app-cons-eq})
are zero. The same is true for the remaining
tables, because most terms in (\ref{app-cons-eq})
vanish because of the integrability conditions. For Tables~\ref{table3} and~\ref{table4} we add
the remark that the condition that the ``weight number'' def\/ined in Table~\ref{table3}
vanishes implies that the third term in equation~(\ref{app-I}) is zero. For
additional details we refer to~\cite{vol-pres}.

\subsection*{Acknowledgement}
  A.W. gratefully acknowledges support from Adam
Krzy\.{z}anowski Fund and Jagiellonian University (grant WRBW
41/07). C.A. and J.S.-G. thank MCyT (Spain) and FEDER
(FPA2005-01963), and support from
 Xunta de Galicia (grant PGIDIT06PXIB296182PR and Conselleria de
Educacion). Further, C.A. acknowledges support from the Austrian
START award project FWF-Y-137-TEC and from the FWF project P161 05
NO 5 of N.J. Mauser.

\pdfbookmark[1]{References}{ref}
\LastPageEnding


\begin{thebibliography}{99}

\footnotesize\itemsep=0pt

\bibitem{AFSG}
Alvarez O., Ferreira L.A., Sanchez-Guillen J.,
A new approach to integrable theories in any dimension,
{\em Nuclear Phys.~B} {\bf 529} (1998), 689--736.

\bibitem{AFZ1}
Aratyn H., Ferreira L.A., Zimerman A.,
Toroidal solitons in (3+1)-dimensional integrable theories,
{\em Phys. Lett. B} {\bf 456} (1999), 162--170.

\bibitem{AFZ2}
Aratyn H., Ferreira L.A., Zimerman A.,
Exact static soliton solutions of (3+1)-dimensional integrable
theory with nonzero Hopf numbers,
{\em Phys. Rev. Lett.}  {\bf 83}  (1999), 1723--1726.

\bibitem{Fer1}
De Carli E., Ferreira L.A.,
 A model for Hopf\/ions on the space-time $S^3\times {\mathbb R}$,
{\em J. Math. Phys.} {\bf 46} (2005), 012703, 10 pages,
\href{http://arxiv.org/abs/hep-th/0406244}{hep-th/0406244}.

\bibitem{Wer1}
Wereszczynski A.,
 Integrability and Hopf solitons in models with explicitly broken
O(3) symmetry,
{\em Eur. Phys. J. C Part. Fields} {\bf 38}  (2004), 261--265,
 \href{http://arxiv.org/abs/hep-th/0405155}{hep-th/0405155}.

\bibitem{Fer2}
Ferreira L.A.,
 Exact time dependent Hopf solitons in 3+1 dimensions,
 {\em J. High Energy Phys.}
{\bf 2006} (2006), no.~3, 075, 9 pages,
\href{http://arxiv.org/abs/hep-th/0601235}{hep-th/0601235}.

\bibitem{Fer3}
Riserio do Bonf\/im A.C., Ferreira L.A.,
 Spinning Hopf solitons on $S^3 \times {\mathbb R}$,
{\em J. High Energy Phys.}
{\bf 2006} (2006), no.~3, 097, 12 pages,
\href{http://arxiv.org/abs/hep-th/0602234}{hep-th/0602234}.

\bibitem{Nic}
Nicole D.A.,
 Solitons with nonvanishing Hopf index,
{\em J. Phys. G}  {\bf 4}  (1978), 1363--1369.

\bibitem{Wer2}
Wereszczynski A.,
 Toroidal solitons in Nicole-type models,
{\em Eur. Phys. J. C Part. Fields}  {\bf 41} (2005), 265--268,
\href{http://arxiv.org/abs/math-ph/0504008}{math-ph/0504008}.

\bibitem{Wer3}
Wereszczynski A.,
 Generalized eikonal knots and new integrable dynamical systems,
{\em Phys. Lett.~B}  {\bf 621} (2005), 201--207,
\href{http://arxiv.org/abs/hep-th/0508121}{hep-th/0508121}.

\bibitem{ASGVW}
Adam C.,   Sanchez-Guillen J.,  Vazquez R.A., Wereszczynski A.,
 Investigation of the Nicole model,
{\em J. Math. Phys.}   {\bf 47} (2006), 052302, 22 pages,
\href{http://arxiv.org/abs/hep-th/0602152}{hep-th/0602152}.

\bibitem{Ward1}
Ward R.S.,
 Hopf solitons on $S^3$ and ${\bf R}^3$,
{\em Nonlinearity} {\bf 12}  (1999), 241--246.

\bibitem{ASGW-S3}
Adam C., Sanchez-Guillen J., Wereszczynski A.,
 Hopf solitons and Hopf $Q$-balls on $S^3$,
{\em Eur. Phys. J. C Part. Fields}  {\bf 47} (2006), 513--524,
\href{http://arxiv.org/abs/hep-th/0602008}{hep-th/0602008}.

\bibitem{Skyrme}
Skyrme T.H.R.,
 A nonlinear f\/ield theory,
{\em Proc. R. Soc. Lond. A} {\bf 260} (1961), 127--138.

\bibitem{Skyrme2}
Skyrme T.H.R.,
 A unif\/ied f\/ield theory of mesons and baryons,
{\em Nuclear Phys.}  {\bf 31} (1962),  556--569.

\bibitem{FSG1}
Ferreira L.A., Sanchez-Guillen J.,
 Inf\/inite symmetries in the Skyrme model,
{\em Phys. Lett. B} {\bf 504}  (2001), 195--200,
\href{http://arxiv.org/abs/hep-th/0010168}{hep-th/0010168}.

\bibitem{BF1}
Babelon O., Ferreira L.A.,
 Integrability and conformal symmetry in higher dimensions:
a model with exact Hopf\/ion solutions,
{\em J. High Energy Phys.}
{\bf 2002} (2002), no.~11, 020, 26 pages.

\bibitem{gen-int}
Adam C., Sanchez-Guillen J.,
 Generalized integrability conditions and target space geometry,
{\em Phys. Lett. B} {\bf 626} (2005), 235--242,
\href{http://arxiv.org/abs/hep-th/0508011}{hep-th/0508011}.

\bibitem{ab-diff}
Adam C., Sanchez-Guillen J., Wereszczynski A.,
 Integrability from an Abelian subgroup of the dif\/feomorphism group,
{\em J. Math. Phys.}  {\bf 47} (2006),  022303, 8 pages,
\href{http://arxiv.org/abs/hep-th/0511277}{hep-th/0511277}.

\bibitem{vol-pres}
Adam C., Sanchez-Guillen J., Wereszczynski A.,
 Conservation laws in Skyrme-type models,
{\em J. Math. Phys.} {\bf 48}  (2007), 032302, 16 pages,
\href{http://arxiv.org/abs/hep-th/0610227}{hep-th/0610227}.

\bibitem{YM-dil}
Adam C., Sanchez-Guillen J., Wereszczynski A.,
 Integrable subsystem of Yang--Mills dilaton theory,
 \mbox{\href{http://arxiv.org/abs/hep-th/0703224}{hep-th/0703224}}.

\bibitem{Schl}
Schlesinger L., Parallelverschiebung und Kr\"ummungstensor,
{\em Math. Ann.}  {\bf 99} (1927), 413--434.

\bibitem{syms}
Adam C., Sanchez-Guillen J.,
Symmetries of generalized soliton models and submodels on target space $S^2$,
{\em J. High Energy Phys.} {\bf 2005} (2005), no.~1, 004, 15 pages,
\href{http://arxiv.org/abs/hep-th/0412028}{hep-th/0412028}.

\end{thebibliography}
\end{document}